# Non-neglectable entropy effect on sintering of supported nanoparticles

Beien Zhu[1, 2†], Shiyuan Chen[3†], Ying Jiang,[4†] Hui Zhang[5†], Rui Qi[2, 6], Yu Han,[2,6] Zhi Liu[5, 7], Bing Yang[8], Matsumoto Hiroaki[9], Chaobin Zeng[9], Wentao Yuan[3], Hangsheng Yang[3], Ze Zhang[3], Jun Hu[1, 2], Yong Wang[3*], and Yi Gao[1, 2*]

[1]Interdisciplinary Research Center, Zhangjiang Laboratory, Shanghai Advanced Research Institute, Chinese Academy of Sciences, Shanghai 201210, China

[2]Key Laboratory of Interfacial Physics and Technology, Shanghai Institute of Applied Physics, Chinese Academy of Sciences, Shanghai 201800, China

[3]State Key Laboratory of Silicon Materials and Center of Electron Microscopy, School of Materials Science and Engineering, Zhejiang University, Hangzhou 310027, China

[4]Materials Genome Institute, Shanghai University, Shanghai 200444, China

[5]State Key Laboratory of Functional Materials for Informatics, Shanghai Institute of Microsystem and Information Technology, Chinese Academy of Sciences, Shanghai 200050, China

[6]University of Chinese Academy of Sciences, Beijing 100049, China

[7]School of Physical Science and Technology, ShanghaiTech University, Shanghai 201210, China

[8]CAS Key Laboratory of Science and Technology on Applied Catalysis, Dalian Institute of Chemical Physics, Chinese Academy of Sciences, 457 Zhongshan Road, Dalian, Liaoning 116023, China

[9]Hitachi High-Tech (Shanghai) Co., Ltd., Shanghai 201203, P. R. China

*Correspondence to: yongwang@zju.edu.cn (Yong Wang); gaoyi@zjlab.org.cn (Yi Gao)

† These authors contributed equally.

**Abstract:** Sintering refers to particle coalescence by heat, which has been known as a thermal phenomenon involving all aspects of natural science for centuries[1]. It is particularly important in heterogeneous catalysis because normally sintering results in deactivation of the catalysts[2-7]. In previous studies, the enthalpy contribution was considered to be dominant in sintering and the entropy effect is generally considered neglectable[8-18]. However, we unambiguously demonstrate in this work that entropy could prevail over the enthalpy contribution to dominate the sintering behavior of supported nanoparticles (NPs) by designed experiments and improved theoretical framework. Using in situ Cs-corrected environmental scanning transmission electron microscopy and synchrotron-based ambient pressure X-ray photoelectron spectroscopy, we observe the unprecedent entropy-driven phenomenon that supported NPs reversibly redisperse upon heating and sinter upon cooling in three systems (Pd-$CeO_2$, Cu-$TiO_2$, Ag-$TiO_2$). We quantitatively show that the configurational entropy of highly dispersed ad-atoms is large enough to reverse their sintering tendency at the elevated temperature. This work reshapes the basic understanding of sintering at the nanoscale and opens the door for various de-novo designs of thermodynamically stable nanocatalysts.

**Main:** Fundamental understanding of sintering is important in surface science, nanoscience, and catalysis. In principle, sintering is considered irreversible because the surface free energy is reduced by bonding small particles to form larger ones. Dispersed smaller particles have larger surface energies and are prone to sintering faster than larger ones. It is well-known that heating always accelerates the particle sintering because it increases the atomic motions, while cooling halts this process. In current theoretical frameworks, only the enthalpy change during a sintering/redispersion process is considered as important and the idea of ignoring entropy contribution is ingrained[4]. Very recently, Goodman et al. observed a high-temperature decomposition of PdO NPs into single atoms on the γ-$Al_2O_3$ support[19]. They still followed the conventional theoretical framework to calculate the free energy by ignoring the direct entropy contribution, although the entropy effect was mentioned. Nevertheless, these conventional theories overlook that when the entropy is taken account into the free energy directly, its contribution can be big enough to surpass the enthalpy contribution at elevated temperatures when the concentration is low. Consequently, the NPs is less stable than the dispersed adatoms and the sintering tendency is completely reversed. For verification, we designed in situ experiments to monitor the sintering and redispersion behaviour of low loading Pd nanoparticles on $CeO_2$ (increasing the configurational entropy) at 200 °C and 500 °C under the same $O_2$ pressure of 10 Pa.

A spherical aberration (Cs-) corrected environmental scanning transmission electron microscopy (ESTEM) was used for the in situ experiments[20]. This state-of-the-art microscope can provide surface sensitive secondary electron (SE) images in real time with a very high spatial resolution under reaction conditions, which is suitable for observing the subtle changes of our sample (Pd-$CeO_2$). The as-prepared NPs have diameters of approximately 1.41 nm, as observed on the $CeO_2$ nanocubes under vacuum (Fig. S1, S2). We first raised the temperature to 500 °C in the $O_2$ environment and took the in situ SE images about 10 minutes later. It showed no visible NP on the $CeO_2$ surfaces, which indicated the supported NPs all redispersed into adatom complexes (Fig. 1a, 1d, Fig. S3a). After cooling the sample for 2 hours, we then took the images at 200 °C and found many sintered NPs (Fig. 1b, 1e, Fig. S3b) with diameters of about 2.05 nm (Fig. S2). These NPs disappeared again (Fig. 1c, 1f, Fig. S3c) when the temperature was raised back to 500 °C. During the experiment, the $O_2$ environment was kept unchanged. In situ ambient pressure X-ray photoelectron spectroscopy (APXPS) characterizations were first performed for the same sample under the same $O_2$ pressure at 500 °C. The Ce *3d*, O *1s*, and the Pd *3d* peaks indicated the full oxidation of the $CeO_2$ surface (Fig. S4)[21] and the Pd sample in the oxygen atmosphere (Fig. 1g). There presented a typical $Pd^{2+}$ in PdO (336.8 eV)[22] and a more ionic state (337.8 eV) that represents the interfacial Pd ions bonding to both O from gas phase and surface lattice O of $CeO_2$, referred as $Pd^{2+}$(PdO/$CeO_2$). We then kept the sample at 200 °C for 1 hour and characterized it at 300 °C to avoid the heavy charging effect at 200 °C[23]. The peak ratio of $Pd^{2+}$(PdO/$CeO_2$) is high at 500 °C correlating to the hot-redispersion and low at 300 °C indicating the cold-sintering. Back to 500 °C, the spectra show much intense peak of $Pd^{2+}$(PdO/$CeO_2$), which is due to the redispersion. No reduction of the sample was found during the temperature control process, meaning that the sintering-dispersion circle shown in Fig. 1a-f was not an oxidation and reduction process[5].

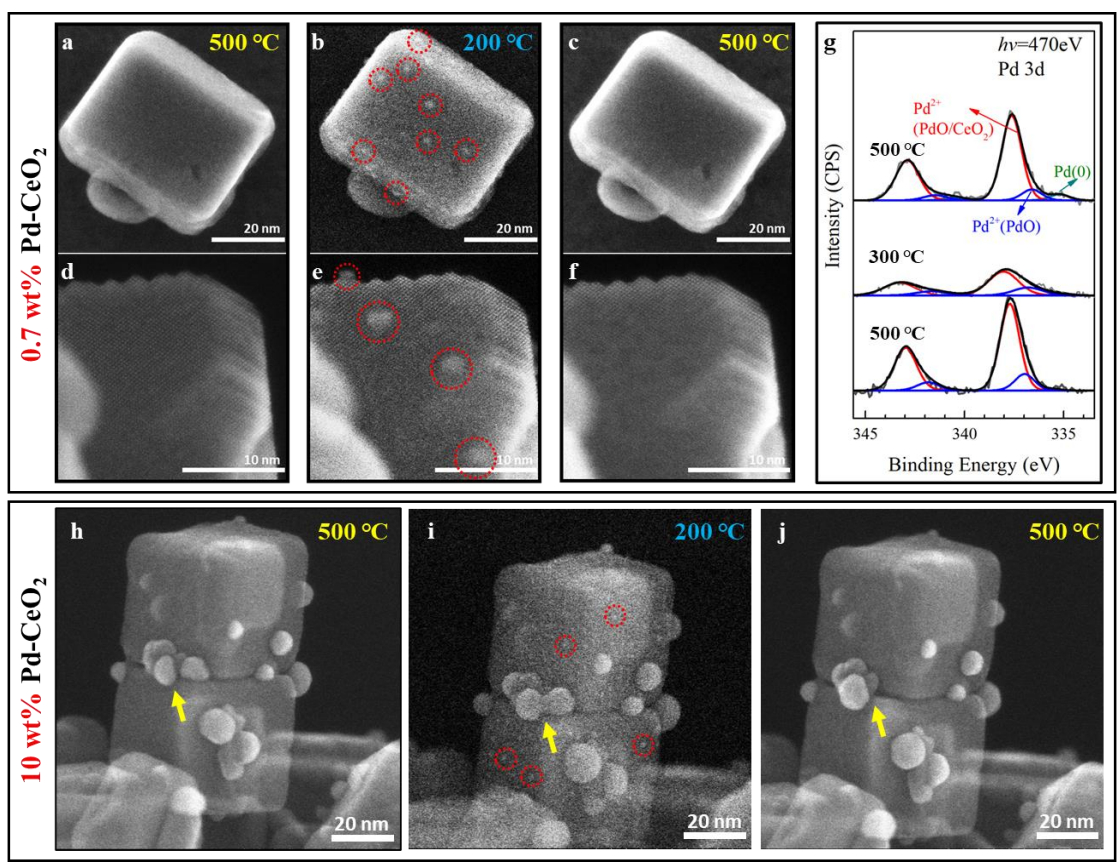

**Figure 1 | ESTEM observations of CeO$_2$ supported PdO NPs in O$_2$ conditions and the corresponding in situ APXPS results. a-c,** The in situ SE images show supported NPs disappeared at 500 °C and appeared at 200 °C. **d-f,** SE images show a similar dispersion-sintering circle in another case. **g,** The APXPS results show that the sample kept oxidized in the O$_2$ environment at different temperatures. **h-j,** In situ SE images of CeO$_2$ supported PdO sample in the O$_2$ conditions with an increased loading of 10 wt%. Some of the reappeared NPs are marked by red circles. The yellow arrows point at where conventional sintering occurs. The pressure of O$_2$ in these experiments was kept at 10 Pa.

We also prepared a Pd-CeO$_2$ sample with a higher loading of 10 wt% (Fig. 1h-j, Fig. S5, S6). After synthesis, both large and small NPs were observed on the CeO$_2$ cubes (Fig. S5, S6a). After raising the temperature to 500 °C for approximately 10 minutes, we observed that all small NPs disappeared in the SE images and the large ones grew larger at their original locations (Fig. 1h, Fig. S6b). The disappearance of the small NPs could be caused by the particle-migration-coalescence or Ostwald ripening.

However, when we cooled the temperature to 200 °C, some small NPs reappeared in the area absent of large NPs, which means their disappearance at 500 °C corresponded to atomic redispersion rather than coalescence or ripening (Fig. 1i, Fig. S6c). Once we raised the temperature to 500 °C, these small NPs disappeared again (Fig. 1j, Fig. S6d). Sintering and redispersion simultaneously occur at the high temperature in the same $O_2$ environment and on the same support facet, which cannot be explained from the perspective of enthalpy.

In recent years, the redispersions of supported metal nanoparticles at high temperatures and in oxygen atmosphere were reported in literature[5, 16, 24-28]. There are two main enthalpy-driven mechanisms to understand the redispersion: 1) surface defects that are generated upon heating trap the dispersed atoms or molecules[13, 24]; 2) the strong interaction between the oxidized metal species and the support surface induce the oxidative redispersion[29]. These redispersed catalysts driven by enthalpy will stay dispersed during the cooling process, which is apparently different from our observations dominated by entropy. In the work of Goodman et al.[19], their calculation of free energy was also based on the enthalpy-dominant framework without involving the entropy contribution. Their entropy effect was attributed to promote the formation of meta-stable single sites, rather than change the relative stability of adatom complexes referenced to the NPs. According to their theory, the complete redispersion of NPs into adatom complexes would not occur since the NPs were anyway thermodynamically more stable. Our observations of the reversible cycle of complete redispersion at the high temperature and resintering at the low temperature obviously show the relative stability between the NPs and the redispersed adatoms is changed at different temperatures, indicating the necessity of revising the free energy calculation by involving the entropy contribution directly.

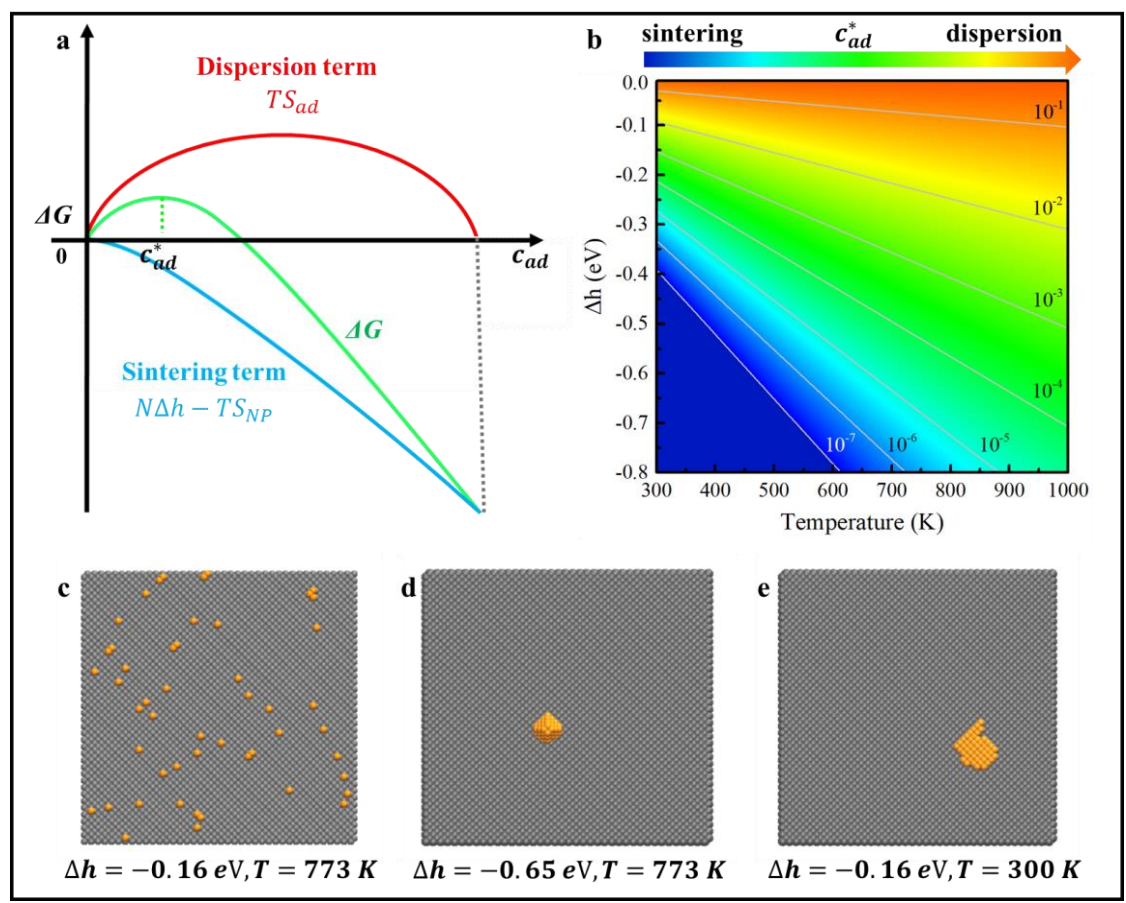

**Figure 2 | Theoretical model and simulations of the entropy effect on the sintering of supported NP. a,** A scheme of the Gibbs free energy difference between the supported nanoparticle and dispersed ad-atoms as a function of surface concentration of ad-atoms. **b,** Contour plots of the critical ad-atom concentration $c_{ad}^*$, where the entropy-driven dispersion most-likely occurs, is plotted as functions of temperature and $\Delta h$. The figure is colour mapped according to the magnitude of $c_{ad}^*$. The larger $c_{ad}^*$, the easier entropy-driven dispersion to occur. The contour lines of $c^*$ from $10^{-1}$ to $10^{-7}$ are marked on the figure. **c-e,** The snapshots of the simulation results represent the entropy effect with different $\Delta h$ and temperature.

To involve the entropy contribution, we consider the Gibbs free energy difference ($\Delta G$) between a supported NP and atomically dispersed ad-atoms (oxidized or reduced) of the NP on the substrate to be described as below:

$$\Delta G = N\Delta h - T(S_{NP} - S_{ad}) \qquad (1),$$

where $N$ is the total number of the ad-atoms. Positive (negative) $\Delta G$ means the dispersed (aggregated) ad-atoms are more stable. $\Delta h$ is the average enthalpy change per ad-atom, that is normally negative. $S_{NP}$ and $S_{ad}$ are the entropies of the NP and ad-atoms, respectively. The vibrational, rotational, translational entropy change between the NP and ad-atoms are neglectable. Following the Boltzmann's entropy formula[30], we considered a coarse-grained model that each atom of the system occupies a lattice site to count the number of possible configurations. Considering that the finite area of the substrate surface is divided into $M$ lattice sites, the Eq. (2) can be rewritten as:

$$\Delta G = [N\Delta h - k_B T ln(C_M^1 e^N)] + k_B T ln C_M^N \qquad (2),$$

where $k_B$ is the Boltzmann constant, $k_B ln(C_M^1 e^N)$ represents the entropy of putting 1 NP in $M$ lattice sites, where the number of local minima of the NP structure is estimated as $e^N$ and each minima is considered to have the same probability to occur[31]. $k_B ln C_M^N$ is the entropy of putting the $N$ ad-atoms in $M$ lattice sites. As shown in Fig. 2a, the first term in Eq. 2 is negative (the sintering term), while the second term is positive (the redispersion term). For large particles, the atoms on the substrate are normally supersaturated. Thus, the redispersion term is neglectable and sintering is meant to occur by heat. However, when the surface concentration of ad-atoms, defined as $c_{ad} = N/M$, is relatively small (Fig. 2a), the ad-atoms can be highly dispersed on surface and the entropy effect could turn $\Delta G$ from negative to positive upon heating (or vice versa upon cooling), resulting in the reversible cold-sintering and hot-redispersion phenomenon (a scheme in Fig. S10). It also explains the simultaneous occurrence of sintering and redispersion in Fig. 1h, j. In the area around a large NP, the local $c_{ad}$ is large and sintering occurs by heating as normal. In the area absent of large NPs, the local $c_{ad}$ is small and heat benefits the redispersion by the entropy effect.

It is interesting to understand under which $c_{ad}$ the entropy effect is non-neglectable. As shown in Fig. 2a, there is a critical $c_{ad}^*$ where the entropy-driven redispersion most-likely occurs, i.e., the one with the highest value of positive $\Delta G$ ($d\Delta G/dN = 0$). To analytically solve this equation, we consider, for large NP, $\Delta h$ is independent of $N$ and the analytic description of $c_{ad}^*$ is:

$$c_{ad}^* = \frac{N^*}{M} = \frac{1}{1+e^{1-\Delta h/k_B T}} \qquad (4).$$

More detailed descriptions of $\Delta h$ and the derivation of $c_{ad}^*$ are shown in the Supplementary Information. The contour plot of $c_{ad}^*$ as a function of $\Delta h$ and $T$ is shown in Fig. 2b. The basic conclusion is that less negative $\Delta h$ and higher $T$ increases $c_{ad}^*$ resulting in more significant entropy effect.

We further performed lattice kinetic Monte Carlo simulations to verify the analytic conclusion for small NPs (details and movies in the Supplementary Information). Two sets of parameters were chosen and the simulated $\Delta h$ was -0.16 eV and -0.65 eV, respectively (Table S1). 50 ad-atoms were first distributed in the 40 × 40 lattice sites of a flat surface ($c_{ad} \cong 0.03$). At 773 K and with $\Delta h = -0.16\ eV$, the dispersed ad-atoms are thermodynamically stable during the 400 million steps simulation (Fig. 2c, Mov. S1), where the analytical $c_{ad}^* = 0.032$. At the same temperature but with $\Delta h = -0.65\ eV$ ($c_{ad}^* = 2.1 \times 10^{-5}$), the ad-atoms sintered to a NP in the beginning stage and kept aggregated in the simulation (Fig. 2d, Mov. S2). The sintering of the ad-atoms also occurred at 300 K with $\Delta h = -0.16\ eV$, where $c_{ad}^* = 7.6 \times 10^{-4}$ (Fig. 2e, Mov. S3). Thus, we confirm that the entropy-driven redispersion prefers higher temperature and lower $\Delta h$. Following these understanding, we observed the cold-sintering and hot-dispersion by in situ ESTEM in two additional systems: Ag-TiO$_2$ and Cu-TiO$_2$. In order to achieve a lower $\Delta h$, the as-synthesized NP size was controlled small, the active TiO$_2$ substrate was chosen, and the experiments were conducted in the O$_2$ environment.

In the experiments, small Ag NPs supported on the anatase TiO$_2$(001) surface were first observed by high-angle anular dark field scanning TEM (HAADF-STEM) under the vacuum at 200 °C (Fig. S7). When the temperature was raised to 550°C under the 10 Pa pressure of O$_2$, some NPs redisappeared (Fig. 3a). Cooling to room temperature and kept for several hours, we observed the cold-sintered NPs, which disappeared again once raising the temperature back to 550 °C (Fig. 3b-c). The sintering-redispersion-sintering circle was also observed in the Cu-TiO$_2$(001) system under the same O$_2$ pressure with a temperature control between 25 °C and 500 °C (Fig. 3d-f). Besides the pure O$_2$ condition, the similar cold-sintering and hot-dispersion of the 0.7 wt% loading Pd-CeO$_2$ catalysts was also found in a CO oxidation condition (Fig. 3g-i, Fig. S8). The mixed gas contained 5% CO and 95% O$_2$ and the total pressure was maintained at 10 Pa. It means the entropy-driven dispersion could occur in real reactions.

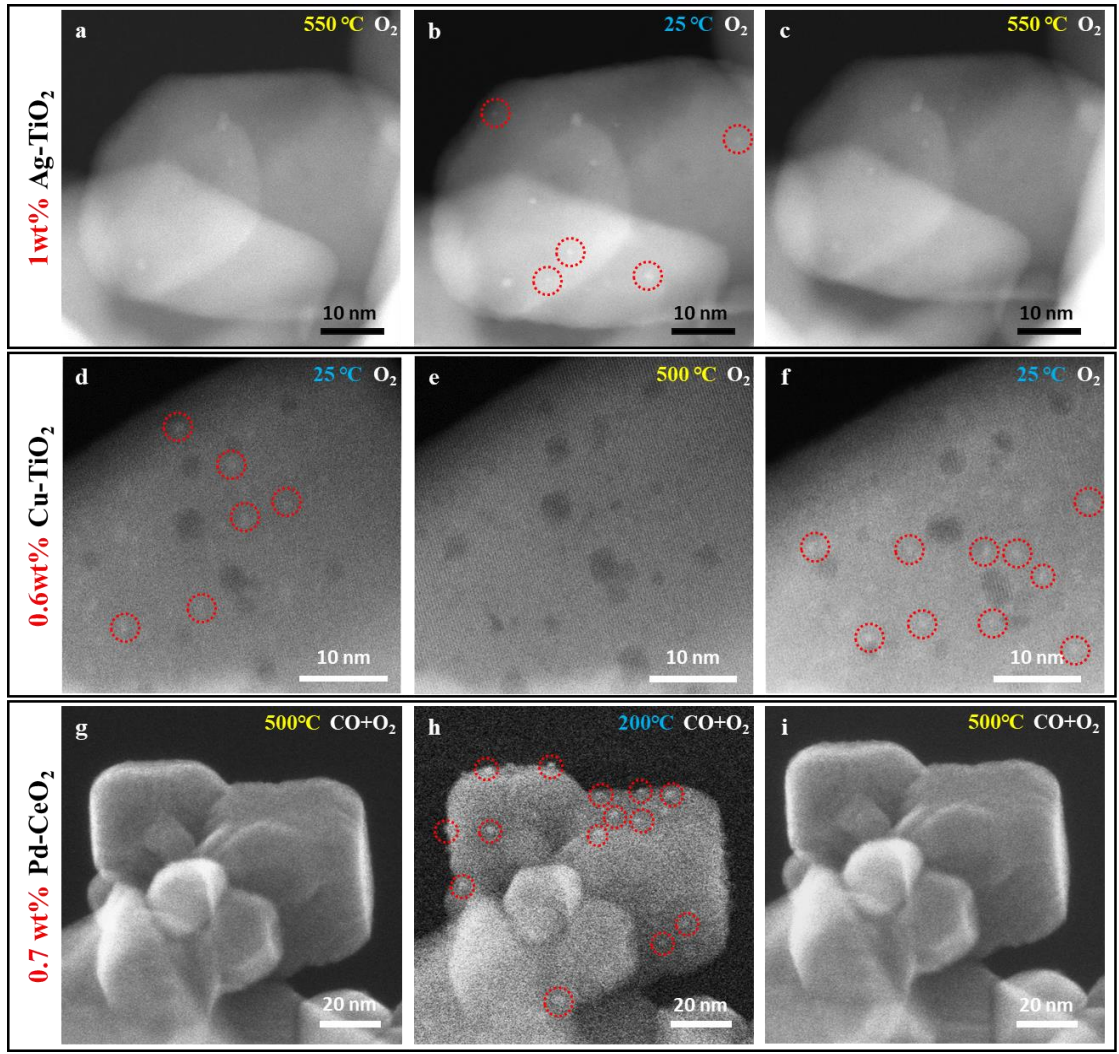

**Figure 3 | In situ ESTEM (a-f: HAADF and g-i: SE) images of various samples under different gas conditions. a-c,** Ag-TiO$_2$ samples and **d-f,** Cu-TiO$_2$ samples in the same O$_2$ environment. **g-i,** Pd-CeO$_2$ samples in the CO and O$_2$ mixed gaseous conditions. The cold-sintering and hot-dispersion were observed in these experiments.

This work highlights the ever-ignored importance of entropy contribution on the sintering behavior in nanoscopic thermodynamics and establishes an applicable theoretical framework of sintering at the nanoscale. From the perspective of entropy, high temeperature can be beneficial rather than harmful to the stability of highly dispersed nanocatalysts, which could eliminate the persistent concern of deactivation of the nanocatalysts resulting from sintering at high reaction temperatures.

**Data availability**

All (other) data needed to evaluate the conclusions in the paper are present in the paper and the Supplementary Materials.

**REFERENCES**

1. "Sinter, v." Oxford English Dictionary Second Edition, *Oxford University Press* (2009).

2. German, R. M. Thermodynamics of sintering. In Sintering of advanced materials (pp. 3-32). Woodhead Publishing (2010).

3. Hansen, T. W., DeLaRiva, A. T., Challa, S. R., & Datye, A. K. Sintering of catalytic nanoparticles: particle migration or Ostwald ripening? *Acc. Chem. Res.* **46**, 1720-1730 (2013).

4. Campbell, C. T. The energetics of supported metal nanoparticles: relationships to sintering rates and catalytic activity. Acc. Chem. Res. 46, 1712–1719 (2013).

5. Morgan, K., Goguet, A., & Hardacre, C. Metal redispersion strategies for recycling of supported metal catalysts: a perspective. *ACS Catal.* **5**, 3430-3445 (2015).

6. Dai, Y., Lu, P., Cao, Z., Campbell, C. T., & Xia, Y. The physical chemistry and materials science behind sinter-resistant catalysts. *Chem. Soc. Rev.* **47**, 4314-4331(2018).

7. Wang, L., Wang, L., Meng, X., & Xiao, F. S. New strategies for the preparation of sinter-resistant metal-nanoparticle-based catalysts. *Adv. Mater.* **31**, 1901905 (2019).

8. Campbell, C. T., Parker, S. C., & Starr, D. E. The effect of size-dependent nanoparticle energetics on catalyst sintering. *Science*, **298**, 811-814 (2002).

9. Joo, S. H. et al. Thermally stable Pt/mesoporous silica core–shell nanocatalysts for high-temperature reactions. *Nat. Mater.* **8**, 126–131 (2009).

10. Lu, J. et al. Coking-and sintering-resistant palladium catalysts achieved through atomic layer deposition. *Science* **335**, 1205–1208 (2012).

11. Parkinson, G. S. et al. Carbon monoxide-induced adatom sintering in a Pd-Fe3O4 model catalyst. *Nat. Mater.* **12**, 724–728 (2013).

12. Ouyang, R., Liu, J. X. & Li, W. X. Atomistic theory of Ostwald ripening and disintegration of supported metal particles under reaction conditions. *J. Am. Chem. Soc.* **135**, 1760–1771 (2013).

13. Jones, J., et al. Thermally stable single-atom platinum-on-ceria catalysts via atom trapping. *Science*, **353**, 150-154 (2016).

14. Zhang, J. et al. Sinter-resistant metal nanoparticle catalysts achieved by immobilization within zeolite crystals via seed-directed growth. *Nat. Catal.* **1**, 540–546 (2018).

15. Yuan, W., et al. Direct In Situ TEM Visualization and Insight into the Facet-Dependent Sintering Behaviors of Gold on TiO$_2$ *Angew. Chem. Int. Ed.*, **57**, 16827–16831 (2018).

16. Xiong, H., et al. Engineering catalyst supports to stabilize PdOx two-dimensional rafts for water-tolerant methane oxidation. *Nat. Catal.* **4,** 830–839 (2021).

17. Yin, P. et al. Quantification of critical particle distance for mitigating catalyst sintering. *Nat. Commun.*, **12**, 1-10 (2021).

18. Hu, S., & Li, W. X. Sabatier principle of metal-support interaction for design of ultrastable metal nanocatalysts. *Science*, **374**, 1360-1365 (2021).

19. Goodman, E.D., Johnston-Peck, A.C., Dietze, E.M. et al. Catalyst deactivation via decomposition into single atoms and the role of metal loading. *Nat. Catal.* **2**, 748–755 (2019).

20. Tang, M., et al. Facet-Dependent Oxidative Strong Metal-Support Interactions of Palladium–TiO2 Determined by In Situ Transmission Electron Microscopy. *Angew. Chem. Int. Ed.*, **60**, 22339–22344 (2021).

21. Kato, S., et al. Quantitative depth profiling of $Ce^{3+}$ in Pt/CeO$_2$ by in situ high-energy XPS in a hydrogen atmosphere. *Phys. Chem. Chem. Phys.* **17**, 5078-5083 (2015).

22. Shafeev, G. A., Themlin, J. M., Bellard, L., Marine, W., & Cros, A. Enhanced adherence of area-selective electroless metal plating on insulators. *J. Vac. Sci. Technol. A.* **14***, 319–326* (1996)*.*

23. Wang, B., et al. Formation and Activity Enhancement of Surface Hydrides by the Metal–Oxide Interface. *Adv. Mater. Interfaces*, **8**, 2002169 (2021).

24. Qu, Y., et al. Direct transformation of bulk copper into copper single sites via emitting and trapping of atoms. *Nat. Catal.*, **1**, 781-786 (2018).

25. Wei, S., et al. Direct observation of noble metal nanoparticles transforming to thermally stable single atoms. *Nat. Nanotech.*, **13**, 856-861 (2018).

26. DeRita, L., et al. Structural evolution of atomically dispersed Pt catalysts dictates reactivity. *Nat. Mater.* **18**, 746–751 (2019).

27. Lin, L., et al. Reversing sintering effect of Ni particles on γ-Mo$_2$N via strong metal support interaction. *Nat. Commun.* **12**, 6978 (2021)

28. Muravev, V., et al. Interface dynamics of Pd–CeO$_2$ single-atom catalysts during CO oxidation. *Nat. Catal.* **4**, 469–478 (2021).

29. Spezzati, G., et al. Atomically dispersed Pd–O species on CeO$_2$ (111) as highly active sites for low-temperature CO oxidation. *ACS Catal.* 7, 6887-6891 (2017).

30. Perrot, P. A to Z of Thermodynamics. *Oxford University Press*. (1998).

31. Wales, D. J., & Doye, J. P. Global optimization by basin-hopping and the lowest energy structures of Lennard-Jones clusters containing up to 110 atoms. *J. Phys. Chem. A*, **101**, 5111-5116 (1997).

## Acknowledgements

We acknowledge the financial support of the National Natural Science Foundation of China (11974195, 52025011, 92045301, 52171019, 51971202, 51872260, 21872145, 21902179). B.Z. thanks for the Youth Innovation Promotion Association, CAS and the financial support of Key Research Program of Frontier Sciences, CAS, Grant NO. ZDBS-LY-7012. Y. W. thanks for the Zhejiang Provincial Natural Science Foundation of China (LD19B030001) and the Key Research and Development Program of Zhejiang Province (2021C01003). The computations were performed at the National Supercomputing Canter in Guangzhou (NSCC-GZ) and Shanghai. The in situ APXPS experiments were performed at BL02B01 of Shanghai Synchrotron Radiation Facility (SSRF).

## Author Contributions

Y. G. initiated the idea. Y. G. and Y. W. supervised the project. B.Z. and Y.G. developed the theory. R. Q., Y. H. performed the calculations and simulations. S. C. synthesized the samples. S. C., Y. J., M. H. and C. Z. conducted the in situ ESTEM experiments. H. Z. conducted the in situ APXPS experiments. Z. L., Y. B., H. Y., W. Y., J. H., and Z. Z. participated in the analysis and discussion. All the authors participated in the preparation of the manuscript.

## Competing interests

The authors declare no competing financial interests.

## Additional information:

**Supplementary information** The online version contains supplementary material
**Correspondence** and requests for materials should be addressed to Y. W. and Y. G.